\documentclass{ws-procs9x6}

\begin{document}

\title{Spacetime Foam, Holographic Principle, and Black Hole Quantum 
Computers}

\author{Y. Jack Ng and H. van Dam}
\address{Institute of Field Physics, Department of Physics and Astronomy,\\
University of North Carolina, Chapel Hill, NC 27599-3255, USA\\
E-mail: yjng@physics.unc.edu}

\maketitle

\abstracts{
Spacetime foam, also known as quantum foam, has its origin in quantum
fluctuations of
spacetime.  Arguably it is the source of the holographic principle,
which severely limits how densely
information can be packed in space.  Its physics is also intimately linked 
to that of black holes and computation.  In particular, the
same underlying physics is shown to govern the computational
power of black hole quantum computers.}

\section{Introduction} 

Early last century, Einstein's general relativity
promoted spacetime from a passive and static arena to an
active and dynamical entity.
Nowadays many physicists also believe
that spacetime, like all matter and energy, undergoes quantum
fluctuations.  These quantum fluctuations make spacetime
foamy on small spacetime scales.  (For a discussion of the relevant
phenomenology and for a more complete 
list of references, see Ref. 1.)

But how large are the fluctuations?
How foamy is spacetime?  Is there any theoretical evidence of
quantum foam?
In what follows, we address these questions.
By analysing a gedanken experiment for spacetime measurement, we show,
in section 2, that spacetime fluctuations scale as the cube root of
distances or time durations.  Then we argue that this cube
root dependence is consistent with the holographic principle.
In section 3, we discuss how quantum foam affects the physics of
clocks (accuracy and lifetime) and computers (computational rate and 
memory space).  We also show that
the physics of
spacetime foam is intimately connected to that of black holes,
giving a poor man's derivation of the Hawking black hole lifetime 
and the area law of black hole entropy.  
Lastly a black hole computer is shown to compute at 
a rate linearly proportional to its mass.

\section{Quantum Fluctuations of Spacetime}
\label{sec:quantum}

If spacetime indeed undergoes quantum
fluctuations, the fluctuations will show up when we measure a distance
(or
a time duration), in the form of uncertainties in the measurement.
Conversely, if in any distance (or time duration) measurement, we cannot
measure the distance (or time duration) precisely, we interpret this
intrinsic limitation to spacetime measurements as resulting from
fluctuations of spacetime.

The question is: does spacetime undergo quantum fluctuations? And
if so, how large are the fluctuations?  To quantify the problem, let us
consider measuring a distance $l$. The question now is: how accurately
can
we measure this distance?  Let us denote by $\delta l$ the accuracy with
which we can measure $l$.
We will also refer to
$\delta l$ as the uncertainty or fluctuation of the distance $l$ for
reasons that will become obvious shortly.  We
will show that $\delta l$ has a lower bound and will use two ways
to calculate it.
Neither method is rigorous, but the fact that the
two very different methods yield the same result bodes well for the
robustness of the conclusion.  (Furthermore, the result is also
consistent with well-known semi-classical black hole physics.  See 
section 3.)

\smallskip
{\bf Gedanken Experiment.}
In the first method, we conduct a thought experiment to measure $l$.
The importance of carrying out spacetime measurements to find the
quantum fluctuations in the fabric of spacetime
cannot be
over-emphasized.  According to general relativity, coordinates do not
have
any intrinsic meaning independent of observations; a coordinate system
is defined only by explicitly carrying out spacetime distance
measurements.
Let us measure the distance between two points.
Following Wigner\cite{wigner}, we put a clock
at one point and a mirror at the other.  Then 
the distance $l$ that we want to measure is
given
by the distance between the clock and the mirror.
By sending a light signal from the clock to the
mirror in a timing experiment, we can determine the distance $l$.
However, quantum uncertainties in the positions of
the clock and the mirror introduce an inaccuracy $\delta l$ in the
distance measurement.  We expect the clock and the mirror to contribute
comparable uncertainties to the measurement.
Let us concentrate on the clock and denote its mass
by $m$.  Wigner argued that if it has a linear
spread $\delta l$ when the light signal leaves the clock, then its position
spread grows to $\delta l + \hbar l (mc \delta l)^{-1}$
when the light signal returns to the clock, with the minimum at
$\delta l = (\hbar l/mc)^{1/2}$.  Hence one concludes that
\begin{equation}
\delta l^2 \gtrsim \frac{\hbar l}{mc}.
\label{sw}
\end{equation}

General relativity provides a complementary bound.
To see this, let the clock be a
light-clock consisting of a spherical cavity of diameter $d$,
surrounded by a mirror wall
of mass $m$, between which bounces a beam of light.  For
the uncertainty in distance measurement not to be greater than $\delta l$,
the clock must
tick off time fast enough that $d/c \lesssim \delta l /c$.  But $d$, the
size of the clock, must be larger than the Schwarzschild radius
$r_S \equiv 2Gm/c^2$ of
the mirror, for otherwise one cannot read the time registered on the
clock.  From these two requirements, it follows that
\begin{equation}
\delta l \gtrsim \frac{Gm}{c^2}.
\label{ngvan}
\end{equation}

The product
of Eq.~(\ref{ngvan}) with Eq.~(\ref{sw}) yields
\begin{equation}
\delta l \gtrsim (l l_P^2)^{1/3} = l_P \left(\frac{l}{l_P}\right)^{1/3},
\label{nvd1}
\end{equation}
where $l_P = (\hbar G/c^3)^{1/2}$ is the Planck length.
(Note that the result is independent of the
mass of the clock and, hence, one would hope,
of the properties of the specific
clock used in the measurement.)
The end result is as simple as it is strange
and appears to be universal: the uncertainty $\delta l$ in
the measurement of the distance $l$ cannot be smaller than the cube root
of $l l_P^2$.\cite{ngvan1}
Obviously the accuracy of the
distance measurement is intrinsically limited by this
amount of uncertainty or
quantum fluctuation.  We conclude that
there is a limit to
the accuracy with which one can measure a distance;
in other words, we can never know the
distance $l$ to a better accuracy than the cube root of $l l_P^2$.
Similarly one can show that we can never know a time duration
$\tau$ to a better accuracy than the cube root of $\tau t_P^2$, where
$t_P \equiv l_P/c$ is the Planck time.
Because the Planck length is so inconceivably short,
the uncertainty or intrinsic limitation to the accuracy in the
measurement
of any distance, though much larger than the Planck length,
is still very small.  For example,
in the measurement of a distance of
one kilometer, the uncertainty in the distance is to an atom as
an atom is to a human being.

\smallskip

{\bf The Holographic Principle.}
Alternatively we can estimate $\delta l$ by applying the holographic
principle.\cite{found,stfoam}
In essence, the holographic principle\cite{wbhts}
says that although the world
around us appears to have three spatial dimensions, its contents can
actually
be encoded on a two-dimensional surface, like a hologram.  To be more
precise,
let us consider a spatial region measuring $l$ by $l$ by $l$.  According
to the
holographic principle, the number of degrees of freedom that this cubic
region
can contain is bounded by the surface area of the region in Planck units,
i.e.,
$l^2 / l_P^2$, instead of by the volume of the region as one may naively
expect.  This principle is counterintuitive, but
is supported by black hole physics in conjunction with
the laws of thermodynamics, and it is embraced by both string theory and
loop quantum gravity.
So strange as it may be, let us now apply the holographic
principle to
deduce the accuracy with which one can measure a distance.

First,
imagine partitioning the big cube into small cubes.
The small cubes so constructed should
be as small as physical laws allow so that we can associate one degree of
freedom with each small cube.   In other words, the number of
degrees of freedom that the region can hold is given by the number of
small
cubes that can be put inside that region.
But how small can such cubes be?
A moment's thought tells us that each side of a small cube
cannot be smaller than the accuracy
$\delta l$ with which we can measure each side $l$ of the big cube.
This can be easily shown by applying the method of contradiction:  assume
that we can construct small cubes each of which has sides less than
$\delta l$.
Then by lining up a row of such small cubes along a side of
the big cube from end to end, and by counting the number of such small
cubes,
we would be able
to measure that side (of length $l$) of the big cube
to a better accuracy than $\delta l$.  But, by
definition, $\delta l$ is the best accuracy with which we can measure
$l$.  The
ensuing contradiction is evaded by the realization that each of the
smallest
cubes (that can be put inside the big cube) measures $\delta l$ by
$\delta l$ by
$\delta l$.  Thus, the number of degrees of freedom in the region
(measuring $l$ by $l$ by $l$) is given by $l^3 / \delta l^3$,
which,
according to the holographic principle, is no more than $l^2 / l_p^2$.
It follows
that $\delta l$  is bounded (from below) by the cube root of $l l_P^2$,
the same result as found
above in the gedanken experiment argument.  Thus, to the extent that the
holographic principle is correct, spacetime indeed fluctuates, forming
foams of
size $\delta l$ on the scale of $l$.  Actually,
considering the fundamental nature of spacetime and the ubiquity of
quantum fluctuations, we should reverse the
argument and then we will come to the conclusion that the
``strange'' holographic principle has its
origin in quantum fluctuations of spacetime.

\section{From Spacetime Foam to Black Hole Computers}
So far there is no experimental evidence of spacetime foam.
In view of this lack of
experimental evidence, we should at least look for theoretical
corroborations (aside
from the ``derivation" of the holographic principle discussed above).
Fortunately such corroborations do exist --- in the sector of black hole
physics.  To
show that, we have to make a small detour to consider clocks and
computers\cite{ng,barrow} first.

\smallskip

{\bf Clocks.}
Consider a clock (technically, a simple and ``elementary'' clock, not
composed of smaller
clocks that can be used to read time separately or sequentially), capable
of resolving time to an accuracy of $t$, for a period of
$T$ (the running time or lifetime of the clock).
Then bounds on the resolution time and the lifetime of the clock can be
derived by following an argument very similar
to that used above in the analysis of the gedanken experiment to measure
distances.
The two arguments are very similar; 
one obtains\cite{ng}
\begin{equation}
t^2 \gtrsim \frac{\hbar T}{mc^2},\hspace{.5in}
t \gtrsim \frac{Gm}{c^3},
\label{clock1}
\end{equation}
the analogs of Eq.~(\ref{sw}) and Eq.~(\ref{ngvan}) respectively.
One can also combine these two equations to give\cite{ng}
\begin{equation}
T/ t^3 \lesssim t_P^{-2} = \frac{c^5}{\hbar G},
\label{clock2}
\end{equation}
the analog of Eq.~(\ref{nvd1}),
which relates clock precision to its lifetime.
(For example, for a femtosecond ($10^{-15}$ sec) precision, the bound on the
lifetime of a simple clock is $10^{34}$ years.)

\smallskip
{\bf Computers.}
We can easily translate
the above relations for clocks into useful relations for a simple
computer
(technically, it refers to a computer designed to perform highly serial
computations, i.e., one that is not divided into subsystems computing in
parallel).
Since the resolution
time $t$ for clocks is the smallest time interval relevant in the
problem, the fastest
possible processing frequency is given by its reciprocal, i.e., $1/t$.
Thus if $\nu$
denotes the clock rate of the computer, i.e., the number of operations
per bit per unit
time, then it is natural to identify $\nu$ with $1/t$.
To identify the number $I$
of bits of information in the memory space of a simple computer, we recall
that the running time $T$ is the longest time interval relevant in the
problem.  Thus,
the maximum number of steps of information processing is given by the
running time divided by the resolution time, i.e., $T/t$.
It follows that one can identify the number $I$ of bits of the
computer with $T/t$.  (One can think of a tape of length $cT$
as the memory space, partitioned into bits each of length $ct$.)
In other words,
the translations from the case of clocks to the case of computers
consist of substituting the clock rate of computation
for the reciprocal of
the resolution time, and substituting the number of bits for the running
time divided by the resolution time.
The bounds on the precision and lifetime of a clock
given by Eq.~(\ref{clock1}) and
Eq.~(\ref{clock2}) are
now translated into bounds on the rate of computation and number of bits
in the computer, yielding respectively
\begin{equation}
I \nu \lesssim \frac{mc^2}{\hbar},\hspace{.3in}
\nu \lesssim \frac{c^3}{Gm},\hspace{.3in}
I \nu^2 \lesssim \frac{c^5}{\hbar G} \sim 10^{86} /sec^2.
\label{computer}
\end{equation}
The first inequality shows that the speed of computation is bounded by
the energy of the computer divided by Planck's constant,
in agreement with the result
found by Margolus and Levitin\cite{ML}, and
by Lloyd\cite{Lloyd} (for the ultimate limits to computation).
The last bound is perhaps even more intriguing: it requires the product
of the
number of bits and the square of the computation rate for
{\it any} simple computer to be less than the square of the
reciprocal of Planck time,\cite{ng}
which depends on relativistic quantum gravity
(involving $c$, $\hbar$, and $G$).  This universal relation links
together our concepts of information/computation, relativity, gravity,
and quantum uncertainty.
Numerically, the
computation bound is about seventy-six orders of magnitude above
what is available for a current lap-top computer performing ten billion
operations per second on ten billion bits, for which
$I \nu^2 \sim 10^{10}/s^2$.

\smallskip
{\bf Black Holes.}
Now we can apply what we have learned about clocks and computers
to black holes.\cite{ng,barrow}
Let us consider using a black hole to measure time.
It is reasonable to use
the light travel time around the black hole's horizon as the resolution
time of the clock,
i.e., $t \sim \frac{Gm}{c^3} \equiv t_{BH}$, then
from the first equation in Eq.~(\ref{clock1}), one immediately finds that
\begin{equation}
T \sim \frac{G^2 m^3}{\hbar c^4} \equiv T_{BH}.
\label{Hawking}
\end{equation}
Thus, if we had
not known of the Hawking lifetime ($T_{BH}$) for
black hole evaporation, this remarkable result
would have implied that there is a maximum lifetime for a black hole!

Finally, let us consider using a black hole to do computations.  This may
sound like a ridiculous proposition.  But if we believe that black holes
evolve according to quantum mechanical laws, it is 
possible\cite{Lloyd}, at least in
principle, to program black holes to perform computations that
can be read out of the fluctuations in the Hawking black hole 
radiation.
How large is the memory space of a black hole computer, and
how fast can it compute?
Applying the results for computation derived above, we readily find
the number of bits in the memory space of a black hole computer, given by
the lifetime of the black hole divided by its resolution time as a clock,
to be
\begin{equation}
I = \frac{T_{BH}}{t_{BH}} \sim \frac{m^2}{m_P^2} \sim \frac{r_S^2}{l_P^2},
\label{bhcomputer1}
\end{equation}
where $m_P = \hbar/(t_P c^2)$ is the Planck mass, $m$ and $r_S^2$ denote the
mass and event
horizon area of the black hole respectively.
This gives the number of bits $I$ as the event horizon area in Planck units,
in agreement with
the identification of a black hole entropy.  Furthermore,
the number of operations per unit time for a
black hole computer is given by
\begin{equation}
I \nu = \frac{T_{BH}}{t_{BH}} \times \frac{1}{t_{BH}} 
\sim \frac{mc^2}{\hbar},
\label{bhcomputer2}
\end{equation}
its energy divided by Planck's constant, as first found by 
Lloyd\cite{Lloyd}.
Note that all the bounds on computation
discussed above are saturated by black hole computers.  Thus one can even
say that once they are programmed to do computations, black holes are the
ultimate simple computers.

\begin{figure}[ht]
\centerline{\epsfxsize=3.3in\epsfbox{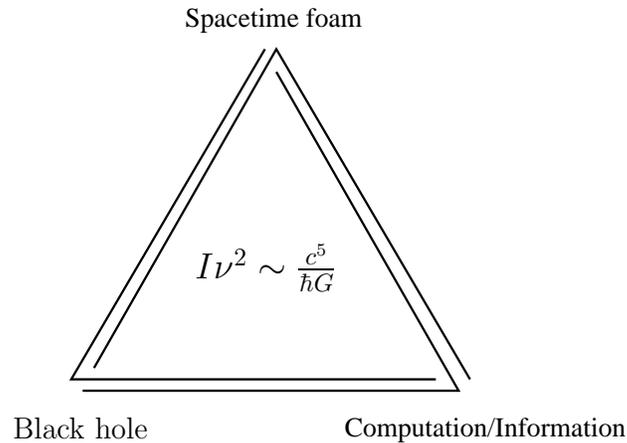}}
\caption{
The {\it universal}
relation for black hole computers, $I \nu^2 \sim  c^5/\hbar G$, 
is a combined
product of the physics behind spacetime foam,
black holes, and computation/information.
\label{fig1}
}
\end{figure}

All these results reinforce the conceptual interconnections of the
physics underlying spacetime foam, black holes, and
computation.
It is intersting that these three subjects
share such intimate bonds and are brought together here 
[see Fig.~\ref{fig1}].
The internal consistency of the physics we have uncovered
also vindicates the simple (some would say overly simple)
arguments we present in section~\ref{sec:quantum},
in the derivation of the limits to spacetime measurements and the elucidation
of the structure of spacetime foam.

\section{Summary}

We have analyzed a gedanken experiment for spacetime
measurements
to show that spacetime fluctuations scale as the cube root of
distances or time durations.  This cube root dependence is
strange, but has been shown to be consistent with the
holographic principle and with semi-classical black hole physics
in general.  (To us, this result for spacetime
fluctuations is as beautiful as it is strange.  Dare we agree with 
Francis Bacon in his observation:
There is no excellent beauty that hath not some strangeness in 
the proportion.)  
We have also shown that
the physics of
spacetime foam is intimately connected to computation.
The unity of physics, underlying spacetime foam,
the holographic principle, black holes, 
and quantum computers, is hereby adequately (if not overwhelmingly) 
demonstrated.

\section*{Acknowledgments}
One of us (YJN) thanks the organizers of the Coral Gables Conference 
for inviting him to present the materials contained in this paper.  
We dedicate
this article to our colleague Paul Frampton on the occasion of his sixtieth
birthday.
This work was supported in part by the US Department of Energy and the
Bahnson Fund of the University of North Carolina.  We thank L.~L. Ng
and T.~Takahashi for their help in the preparation of this manuscript.

\end{document}